\def\be{\begin{equation}}
\def\ee{\end{equation}}
\def\bea{\begin{eqnarray}}
\def\eea{\end{eqnarray}}
\documentstyle[preprint,aps,eqsecnum]{revtex}
\begin{document}
\preprint{April, 1994}
\title
{Towards exact bosonization
\\
of the Calogero-Sutherland model}

\author{D.V.Khveshchenko}
\address
{Physics Department, Princeton University,
\\
Princeton, NJ 08544\\
and
Landau Institute for Theoretical Physics,
\\
2, st.Kosygina, 117940, Moscow, Russia}
\maketitle

\begin{abstract}
\noindent
We demonstrate that the exact form-factors of the Calogero-Sutherland model
which were recently found in \cite{Ha} in confirmation of the congectures
earlier made by Haldane (Ref.\cite{Hal1}) can be reproduced
in the framework of some bosonization procedure
in momentum space. This observation
implies a possibility of an exact bosonization of this model describing
one-dimensional anyons.
\end{abstract}
\pagebreak

The Calogero-Sutherland (CS) model \cite{CS}
of one-dimensional (1D) spinless
particles with long-ranged
$1/x^2$ interaction
( $V(x)={\pi^2\lambda(\lambda-1)
\over {L^2(\sin {\pi x\over L})^2}}$ in a finite size $L$
system) provides a remarkable example of an exactly solvable system
which can be viewed as a continuous deformation away from either
ideal Fermi or Bose gas.

At special values $\lambda=1/2, 1$ and $2$ the original continuous
model describes
statistics of eigenvalues of random matrices belonging to one of the three
Dyson's ensembles. Recently these special cases  were intensively studied
in the framework of the theory of universal spectral
correlations in random systems
\cite{SAL}.
The lattice version of the CS model at $\lambda=2$ also
induced a formulation of the Haldane-Shastry spin chain model \cite{HS}
possessing rich Yangian symmetry in the case of a finite length chain
\cite{HHTBP}.
Further generalizations introducing particles
with internal degrees of freedom in both continuous and lattice versions
led to a family of new solvable models with graded $SU(m|n)$
symmetries \cite{HH}.

The spectrum of the CS
model obtainable  by means of the so-called Asymptotic
Bethe Ansatz (ABA) \cite{S} appears to be  strikingly simple and
enables one to consider this system as an ideal gas of excitations
obeying fractional exclusion principle formulated in \cite{Hal2}. The latter
was originally proposed as an alternative definition of
fractional
quantum statistics which is independent of the space dimension in contrast
to the conventional notion of exchange statistics based on braiding relations
for corresponding operators \cite{?}.

Remarkably, in the framework of the CS model
these two definitions of fractional statistics demonstrate their
 perfect consistency.
In its fractional exclusion formulation
the principle of fractional statistics
manifests itself as a set of rules governing allowed patterns
of occupation numbers of states with
momenta $k_i$ obeying ABA equations \cite{S} for a system
with  periodic boundary conditions
\be
Lk_i=2\pi n_i +\pi(\lambda-1)\sum_{j\neq i}sgn(k_i-k_j)
\ee
Eigenvalues of the Hamiltonian and the total momentum operator
are given by free particle's expressions: $E={1\over 2m}\sum_i k^{2}_i$ and
$P=\sum_i k_i$. In the case of a rational coupling constant
$\lambda={M\over N}$ the ground state
$(z_i=e^{{2\pi ix_i\over L}})$
\be
\Psi_{0} (\{z_i\})=\prod_{i<j}^{\cal N}(z_i-z_j)^\lambda
\prod_{i}^{\cal N}z_i^{-\lambda{{\cal N}-1\over 2}}
\ee
can be characterised as a periodic
array of one occupied and $M-1$ empty states associated with  consequent
non-negative integers $n_i$.
Patterns describing excited states are
also subjected to a very restrictive selection
rule which dictates that one can only move $N$ momenta out of the ground state
distribution simultaneously. This rule can be naturally interpreted in terms
of fundamental
quasiparticles with disperion $\epsilon_{q}(u)={1\over 2}(u^2-k_{F}^2)$,
($|u|>k_{F}=\pi\rho$ where $\rho$ is a density of particles
of mass $m=1$)  and quasiholes
($\epsilon_{h}(v)={1\over 2\lambda}(k_{F}^2-v^2)$, $|v|<k_{F}$)
as a statement that  an elementary
excitation at conserved particle number is constituted by $N$ quasiparticles
and $M$ quasiholes. On the basis of this constraint
it was first anticipated in \cite{Hal1} and then explicitly found in
\cite{Ha} that
dynamical correlation functions
of local operators such as
a fundamental field operator $\Psi(x)$ and density $\rho(x)$
have exclusively simple structure. It stems from the fact
that sums over intermediate states $|\nu>$ in formal expansions
(where $O=\Psi,\rho$ )
\be
<vac|O(x,t)O(0,0)|vac>=\sum_{\nu} |<\nu|O(0,0)|vac>|^2
\exp(iE_{\nu}t-iP_{\nu}x)
\ee
now amount to single terms
given by the only nonzero form-factors
\be
<u_1,...u_{N-1};v_1,...v_M |\Psi(0)|vac>=C_{M,N-1}
{\prod^{M}_{i<j}(v_i-v_j)^{1/\lambda}
\prod^{N-1}_{i<j}(u_i-u_j)^{\lambda}
\over {\prod_{i}^{M}(1-v^{2}_i)^{1-1/\lambda\over 2}
\prod_{i}^{N-1}(u^{2}_i-1)^{1-\lambda\over 2}
\prod^{M(N-1)}_{i,j}(u_i-v_j)}}
\ee
and
\be
<u_1,...u_N; v_1,...v_M |\rho(0)|vac>=
C_{M,N}{\prod^{M}_{i<j}(v_i-v_j)^{1/\lambda}
\prod^{N}_{i<j}(u_i-u_j)^{\lambda}
(\sum_{i}^{N}u_i-{1\over \lambda}\sum_{i}^{M}v_i)\over
{
\prod_{i}^{N}(u^{2}_i-1)^{1-\lambda\over 2}
\prod_{i}^{N}(1-v^{2}_i)^{1-1/\lambda\over 2}\prod^{NM}_{i,j}(u_i-v_j)}}
\ee
where $C_{nm}$ are known normalization factors \cite{Hal1},\cite{Ha} and
rapidities are normalized with respect to the Fermi momentum ($
k_i\rightarrow k_i/k_F$).

Formulae (4) for $\lambda=2$ \cite{HZ} and (5) for $\lambda=2,1$
and $1/2$ \cite{LSA}
were first obtained by means  of supermatrix model calculus.
In \cite{Ha} they were found at arbitrary rational $\lambda$ by using
the method of Jack polynomials which was  developed in \cite{F}
to compute static correlations functions related to Selberg integrals.

This basically combinatoric approach explicitly implements
the fractional exclusion principle. On the other hand, in view
of the abovementioned equivalence of two definitions of fractional statistics
one might expect that the problem also allows a purely algebraic solution
based on monodromy properties of quasiparticle field operators and their
braiding relations.

Indeed, one might be tempted to interpret formulae (4)and (5) as
vacuum expectation values of some "vertex
operators" constructed from a free boson field (
2D Coulomb gas correlators).
In this  Letter we  present arguments in favor
of such an interpretation which should be actually viewed as a possibility
to carry out an exact bosonization of the CS model.

In its original
formulation the conventional bosonization procedure
was developed to describe longwavelength properties of gapless 1D fermions
with contact interactions
in terms of their collective excitations (charge and spin
density waves). The procedure involves a coordinate space
construction of the
fundamental fermion field operator
in terms of bosons ($\psi(x)\sim\sum_{R,L}e^{ i\phi_{R,L}(x)\pm ik_{F}x}$)
which is suitable only for a description of low-energy
fermionic excitations with linear dispersion in the vicinity of
well separated Fermi points.
Apparently, intending to develop an exact
 bosonization scheme valid at all scales
one can no longer
use this simple representation.

Conceivably, some hints for
a proper construction in coordinate space can be inferred from the fact
that the wave function of the ground state
(2) (as well as lowest excited states) can be reproduced as vacuum
expectation values of
a product of $\cal N$ "quasi-chiral" vertex operators
$\psi(z)=e^{i{\sqrt \lambda}\phi(z)}$
written in terms of some formal bosonic variable
\be
\Psi_{0}(\{z_i\})=<\prod_{i}^{\cal N}e^{i{\sqrt\lambda}\phi(z_i)}
e^{-i{\lambda\over 2}({\cal N}-1)\phi(0)}>
\ee
where the gaussian average has to be taken as
$<\phi(z)\phi(z^{\prime})>=-\ln(z-z^{\prime})$.
Analogously, the wave function
$\Psi(z)=\Psi_{0}(z)\prod_{i}^{\cal N}\prod_{j}^{{\cal N}_h}(z_i -\zeta_j)$
corresponding to
the state with  ${\cal N}_h$ quasiholes
can be readily obtained by inserting a product of ${\cal N}_h$ operators
$e^{{i\over {\sqrt \lambda}}\phi(\zeta)}$
into (6).

These observation are equivalent to
those made about Laughlin trial wave functions \cite{MR} which are
nothing but a result of analytic continuation of the
CS model wave functions defined on a unit circle
into an entire complex  $z$ plane (if nonanalytic gaussian
factors $e^{-{|z_i|^2/ 4}}$ are, as usual,
asigned to integration measure).

To make this heuristic construction physically meaningful we
implement another observation from the so-called collective field
theory of unitary matrix models
\cite{J}. It was shown in \cite{J} that the system of
free nonrelativistic fermions is equivalent to
the cubic Hamiltonian
\be
H={\rho\over 2}((\partial\varphi)^2+(\partial\theta)^2)
+{1\over 2}(\partial\varphi)^2\partial\theta
+{1\over 6}(\partial\varphi)^3
\ee
where bosonic variables $\varphi(x)$ and $\theta(x)$ are conjugated
to each other:
$
[\varphi(x),\theta(y)]={i\pi\over 2} sgn(x-y)
$ and related to local particle number $(\rho={1\over \pi}\partial \varphi)$
and
current $(j={1\over \pi}\partial \theta)$ densities.
Elaborating on this theory one can see
that although time evolution of bosonic variables
is quite complicated,
it turns out that some exponential
composite operators $W_{ab}=\exp(ia \varphi+ib\theta)$ may
have much simpler dynamics.
In particular, the exponentials
$
W_{1,1}=\exp(i\varphi+i\theta)
$
satisfies the equation of motion for free fermions.

Moreover the exponentials
$W_{\lambda,1}=\exp(i\lambda\varphi+i\theta)$
obey the equation of motion of the CS model at arbitrary
$\lambda$ (modulo some normal ordering ambiguities). We believe
that this circumstance
makes it possible to consider $W_{\lambda,1}$
as a plausible representation
of the fundamental
CS field with time dependence resulting from
the evolution of variables  $\varphi(x)$ and $\theta(x)$ determined by the
cubic Hamiltonian (7). It implies that
 the "quasi-chiral" boson field appearing in (6)
can be identified as $\phi(z)=\lambda\varphi+\theta$.

Referring to $W_{1,1}$ as a  bare
fermion $\psi(x)$
one can also treat $W_{\lambda,1}$ as a "dressed" anyon field operator:
\be
{\tilde \psi}(x)
=e^{i(\lambda-1)\pi\int^{x}_{-\infty}\psi^{\dagger}(y)\psi(y)dy}
\psi(x)
\ee
which is another formulation  of the Eq.(1).
In terms of ${\tilde \psi}(x)$
 the CS Hamiltonian becomes quadratic which
sheds additional light on the remarkable simplicity of the
spectrum of the CS model describing purely statistical interaction.

However this real space bosonization approach doesn't provide an immediate
construction of eigenstates of the model. Even for the case
$\lambda=1$ studied in the context of the $1+1$-dimensional string theory
in \cite{AJ} it appears to be
quite involved.

On the other hand, this aim can be readily achieved in the framework of a
dual approach of "momentum space bosonization"
recently proposed in \cite{LL}. Being applied to the Sine-Gordon model at
$\beta=1$ this method
 provided a bosonic representation of a massive relativistic
fermion field which facilitates computations of formfactors of other
operators built from the Sine-Gordon field.
This study also clarified a physical origin of the bosonic field
$\phi(z)$ from (6) as a generating
function for the integrals of motion $I_n =\sum_{i}p_{i}^{n}$
of the free fermion system
\be
\phi(z)=\sum_{n\neq 0}{I_n\over in}
\ee
where $I_n$ at $n<0$ are "angle" variables conjugated to those at $n>0$
("action" variables). Altogether they form a $U(1)$ Kac-Moody algebra:
  $[I_n,I_m]=n\delta_{nm}$.

Being formulated this way
the method of \cite{LL}
allows a straightforward generalization onto the case of
interacting systems which are characterised by
factorizable scattering and possess $\cal N$ integrals of motion
given by powers of some Lax operator: $I_n =Tr L^n$.
Since the CS model determined by the  Lax operator
$L_{ij}=\delta_{ij}p_{i}+{\pi\lambda
\over L}(1-\delta_{ij}){\cot {\pi\over L}(
x_i -x_j)}$
allows the ABA solution, the eigenvalues
of integrals of motion can be also presented  in the
free particle form $I_n |\{k_i\}>=\sum_{i}k_{i}^{n}|\{k_i\}>$
 where $k_i$ are asymptotic momenta obtainable from (6).

In the picture of second quantization
the integrals of motion acquire free particle form in terms of
quasiparticle and quasihole creation-annihilation
 operators $Z_a (u)$ $(a=p,h)$ labeled by
asymptotic momenta. Being considered
on general
grounds operators $Z_a$ obey
Zamolodchikov-Faddeev (ZF) algebra $(a=1,...,f)$ \cite{ZF}:
\bea
Z_a (u)Z_b (v)=S_{ab}(u-v)Z_b (v)Z_a (u)\nonumber\\
Z^{\dagger}_a (u)Z^{\dagger}_b (v)=S_{ab}
(u-v)Z^{\dagger}_b (v)Z^{\dagger}_a (u)\nonumber\\
Z_a (u)Z^{\dagger}_b (v)=S_{ab}^{-1}
(u-v)Z^{\dagger}_b (v)Z_a (u)+\delta_{ab}\delta(u-v)
\eea
where $S_{ab}(u-v)$ is a scattering matrix of excitations
which depends on
relative rapidity.
In the case $S_{ab}(u)=\gamma_a\gamma_b S(u)$
($\gamma_a$ is a $U(1)$ "charge" of a species $a$)
bosonization procedure can be performed by constructing a
representation of the algebra (10)
$Z_a (u)=e^{i\gamma_a \Phi(u)}$
in terms of a one-component bosonic field with commutation relations
\be
[\Phi(u),\Phi(v)]=\ln S{(u-v)}
\ee
which infer the vacuum expectation value
\be
<0|\Phi(u)\Phi(v)|0>=-{\ln f(u-v)}
\ee
with $f(u)$ determined by the equation $S(u)=f(-u)/f(u)$.
In terms of ZF operators the integrals of motion
can be written as
\be
I_n =\sum_{a}^{f}\int_{\Lambda_a}
 du Z_a ^{\dagger}(u)p^{n}_a (u) Z_a (u)
\ee
where $\Lambda_a$ is a  region of rapidities supporting excitations $Z_a$
with momenta $p_a (u)$.

Expressions for
local (in momentum space) operators bilinear in $Z_a$ can be found in a manner
similar to the point-splitting procedure in the conventional real space
bosonization which yields
\be
Z_a ^{\dagger}(u)Z_a(u)={\gamma_a\over \pi} {\partial\Phi(u)
\over \partial u}
\ee
 With the use of (14)  the bosonic phase (9) providing a real space
representation of the fundamental field operator $\psi(z)$ can be recovered
in the form
\be
\phi(z)=\sum_{a}^{f}{\gamma_a\over 2}\int_{\Gamma_a}{du\over 2\pi i }
\ln(1-u^2 z^2){\partial\Phi(u)\over \partial u}
\ee
where integrals in the complex plane of
rapidities are taken over contours $\Gamma_a$  encircling real axis intervals
$\Lambda_a$.

In the particular case of
the  CS model scattering matrix is given by matrix elements
$S_{pp}(u_1 -u_2)=\pi\lambda sgn(u_1 -u_2)$,
$S_{hh}(v_1 -v_2)={\pi\over \lambda} sgn(v_1 -v_2)$ and
$S_{ph}(u -v)={\pi}sgn(u-v)$ equal to statistical phase factors assigned
to elementary excitations having $\gamma_{p,h}=\pm\lambda^{\pm 1/2}$
with rapidities equal to their velocities.
It then leads to ZF operators
\be
Z_{p}(u)=e^{i\sqrt\lambda\Phi(u)},~~~~~~
Z_{h}(v)=e^{ -i{\Phi(v)\over \sqrt\lambda}}
\ee
creating eigenstates $(|u_i;v_j>=\prod_{i}Z_{p}^{\dagger}(u_i)
\prod_{j}Z_{h}^{\dagger}(v_j)|vac>)$ and $<0|\Phi(u)\Phi(v)|0>=-\ln (u-v)$.
By construction $\Phi(u)$ is subjected to
the condition $\Phi(0)=0$ and has no singularities on the real axis.
 Naturally, $\Phi(u)$ should also vanish at
$|u|\rightarrow \infty$.
Although general rules of choosing integration
contours in (15) remain to be clarified we believe
that a correct
prescription
is to take contour intergals at positive and negative Re$u$ separately
to keep a distinction between left and right moving excitations.

Then integrals taken over four contours encircling intervals $(-\infty,-1],
[-1,0), (0,1],$ and $[1,\infty)$
yield the following expression for the fundamental CS field
operator  at $z=1$ $(x=0)$:
\be
\psi(0)\sim\exp{i\over 2}({\sqrt \lambda}-{1\over {\sqrt \lambda}})(\Phi(1)+
\Phi(-1))
\ee
Together with the formulae for $Z_{p}(u)$ and $Z_{h}(v)$
given by (16) the gaussian average
\be
<u_1,...u_{N-1}; v_1,...v_M |\psi(0)|vac>=
<\prod_{i}^{N-1}Z_{p}(u_i)\prod_{j}^{M}
Z_{h}(v_j)\psi(0)>
\ee
 reproduces the form-
factor
(4) up to a normalization constant.
The abovementioned selection rules
constraining allowed quasiparticle contents of asymptotic states
provide that (18) is free of infrared divergent factors potentially
coming out of  averages
$<\Phi^{2}(u)>$ at coincident momenta.

Notice that one  can only take the free fermion
limit $(M=N=1)$ in the final formula (4)
to obtain $<v|\psi(0)|vac>=\theta (1-|v|)$.

By inspection it is also easy to
see that to reproduce the form-factor of the density
operator
(5) one can use
the following prescription
\be
\rho(0)\sim P
 \exp{i\over 2}({\sqrt \lambda}-{1\over {\sqrt \lambda}})(\Phi(1)+\Phi(-1))
\ee
where $P=\sum_{i}^{N}u_i -{1\over \lambda}\sum_{j}^{M}v_j$
is a total momentum of excitations which can be identified with
$(\partial_z \phi(z)-{\bar \partial}_{\bar z} \phi({\bar z}))|_
{z\rightarrow 1}$.
The relation (19) should result from an operator algebra of
$\psi(z)$ and $\psi^{\dagger}(z)$ at arbitrary $z$
which remains to be established.

In conclusion  we notice that a systematic approach to
a complete bosonization of the CS model could be built on the basis
 of
its infinite symmetry recognised as $W_\infty$ \cite{HW}.
 The corresponding  linear algebra is formed by operators
$W^{s}_{n}=\sum_{i}^{\cal N}x_{i}^{s-1}p_{j}^{n+s-1}+...
$ which can be recursively obtained as
$W^{s}_{n}={1\over 2(n+s)}[\sum_{i}^{\cal N}x_{i}^{2}, W^{s-1}_{n+2}]$
from the CS
integrals of motion: $W^{1}_{n}=I_n$.
In particular, the CS Hamiltonian is among these operators
($H_{CS}=W^{1}_{2}$). Representation theory
of $W_\infty$ in terms of free bosons discussed in
\cite{W} might appear to be useful here.

A complete CS model bosonization scheme valid at all scales
would be an important step towards an exact formulation of a
(nonlinear) hydrodynamics of 1D fluids of interacting fermions.

The author is indebted to F.D.M.Haldane for valuable discussions and
for the opportunity to
become aware of Refs.\cite{Hal1} and \cite{Ha} prior to publication.
This work was supported by the NSF Grant.

\end{document}